# STANDARD MODEL PARAMETERS FROM QUARKONIA USING LATTICE QCD

AIDA X. EL-KHADRA*

*Physics Department, Ohio State University, 174 W 18th Ave*
*Columbus, OH 43210*
*E-mail: aida@pacific.mps.ohio-state.edu*

Quarkonia – mesons made of a heavy quark and anti-quark – have been extensively studied experimentally. Theoretical calculations of quarkonia based on lattice QCD are possible with control over the systematic errors. The comparison with experimental measurements of the quarkonia spectra leads to determinations of Standard Model parameters: the strong coupling, $\alpha_s$, and the heavy quark masses.

## 1. Introduction and Motivation

By now, we have accumulated a large body of circumstantial evidence that QCD is the correct theory of the strong interactions. What is sorely missing is a first-principles understanding of the non-perturbative effects QCD gives rise to, the most dramatic of them being the observed hadron spectrum. On the practical side, this lack of understanding limits the extraction of Standard Model parameters from experimental measurements. Lattice field theory offers a systematic first principles approach to solving QCD.

Quarkonia are at present the best understood hadronic systems. Both the charm and bottom quark masses are large compared to the typical QCD scale, $\Lambda_{QCD}$. The $b\bar{b}$ and $c\bar{c}$ bound states are therefore governed by non-relativistic dynamics. While the QCD potential was not known from first principles, relatively simple guesses for phenomenological potentials have proven quite successful in describing the experimentally measured bound state spectra of quarkonia[1]. As has been argued by Lepage[2], quarkonia are also the easiest systems to study with lattice QCD, with the potential of leading to a complete first-principles understanding of this simple system.

Finite-volume errors are much easier to control for quarkonia than for light hadrons. Lattice-spacing errors, on the other hand, can be larger for quarkonia and need to be considered. An alternative to reducing the lattice spacing in order to control this systematic error is improving the action (and operators). For quarkonia, the size of lattice-spacing errors in a numerical simulation can be *anticipated* by calculating expectation values of the corresponding operators using potential model wave functions. They are therefore ideal systems to test and establish improvement techniques.

Most of the work of phenomenological relevance is done in what is generally re-ferred to as the "quenched" (and sometimes as the "valence") approximation. In this approximation gluons are not allowed to split into quark - anti-quark pairs (sea quarks). In the case of quarkonia, potential model phenomenology can be used to estimate this systematic error.

---

*adress after Sept. 1 1995: Physics Department, University of Illinois, 1110 W. Green St., Urbana, IL 61801

Control over systematic errors in turn allows the extraction of Standard Model parameters from the quarkonia spectra.

The rest of this talk proceeds as follows: Section 2 gives a brief introduction to lattice QCD methods without going into technical details, since a number of pedagogical introductions and reviews already exist in the literature[3]. The following sections 3-5 discuss quarkonium results and determinations of Standard Model parameters. Section 6 finally concludes with some remarks about future prospects.

## 2. An Introduction to Lattice QCD

Starting with the Feynman path integral formulation in Euclidean space, the discretization of space-time (with lattice spacing $a$) regulates the integral at short distances or in the ultraviolet. A finite volume (of length $L$) is necessary for numerical techniques and also introduces an infrared cut-off or momentum-space discretization. The vacuum expectation of a Greens function, $\mathcal{G}$, which is a product of gauge and fermion fields, is defined as:

$$\langle \mathcal{G} \rangle = \lim_{L \to \infty} \lim_{a \to 0} \langle \mathcal{G} \rangle_{L,a} \;, \quad \langle \mathcal{G} \rangle_{L,a} = Z_{L,a}^{-1} \int \mathcal{D}\psi \mathcal{D}\bar{\psi} \mathcal{D}U \, \mathcal{G} \, e^{-S} \;. \tag{1}$$

$Z_{L,a}$ normalizes the expectation value. I have omitted spin and color indices for compactness. The gauge degrees of freedom are written as (path ordered) exponentials of the gauge field, $A_\mu$:

$$U_\mu(x) = e^{i \int_x^{x+a} dx' A_\mu(x')} \simeq e^{ia A_\mu(x)} \;, \tag{2}$$

which makes it easy to maintain gauge invariance. The link fields, $U$, are $SU(3)$ matrices. The (Euclidean) QCD action,

$$S = S_g + S_f \;, \quad S_g = \frac{1}{4g^2} \int d^4x \, F_{\mu\nu} F^{\mu\nu} \;, \quad S_f = \int d^4x \, \bar{\psi}(x)(\slashed{D}+m)\psi(x) \;. \tag{3}$$

is discretized, such that Eq. (3) is recovered in the the continuum ($a \to 0$) limit:

$$S_{\text{lat}} = S + \mathcal{O}(a^n) \;, \quad n \geq 1 \;. \tag{4}$$

I will not go into the explicit formulations of $S_{\text{lat}}$ here, but instead refer the reader to pedagogical introductions [3]. The most common form for the gauge action is Wilson's[4], written in terms of plaquettes – products of $U$ fields around the smallest closed loop on a lattice. Wilson's gauge action has discretization errors of $\mathcal{O}(a^2)$.

For fermions the situation is more complicated. The discretization of

$$M \equiv \slashed{D} + m \;,\qquad(5)$$

is a sparse, finite dimensional matrix. Two different approaches are in use. In Wilson's formulation[5] chiral symmetry is explicitly broken, but restored in the continuum limit. The pay-off is a solution of the so-called fermion doubling problem. Staggered fermions[6] keep a $U(1)$ chiral symmetry at the expense of dealing with 4 degenerate flavors of fermions.

Eq. (1) emphasizes that QCD is a limit of lattice QCD. However, in numerical calculations these limits cannot be taken explicitly, only by extrapolation. This is feasible, because theoretical guidance for both limits is available. The zero-lattice-spacing limit is guided by asymptotic freedom, since the lattice spacing is related to the gauge coupling by the renormalization group. Quantum field theories in large but finite volumes have also been analyzed theoretically[7].

In a numerical calculation the limits are taken by considering a series of lattices, as illustrated in Figure 1. While keeping the physical volume (or $L$) fixed, the lattice spacing is successively reduced; then, keeping the lattice spacing fixed the volume is increased. The calculation is in the continuum (infinite volume) limit once the hadron spectrum or matrix elements of interest become independent of the lattice spacing (volume).

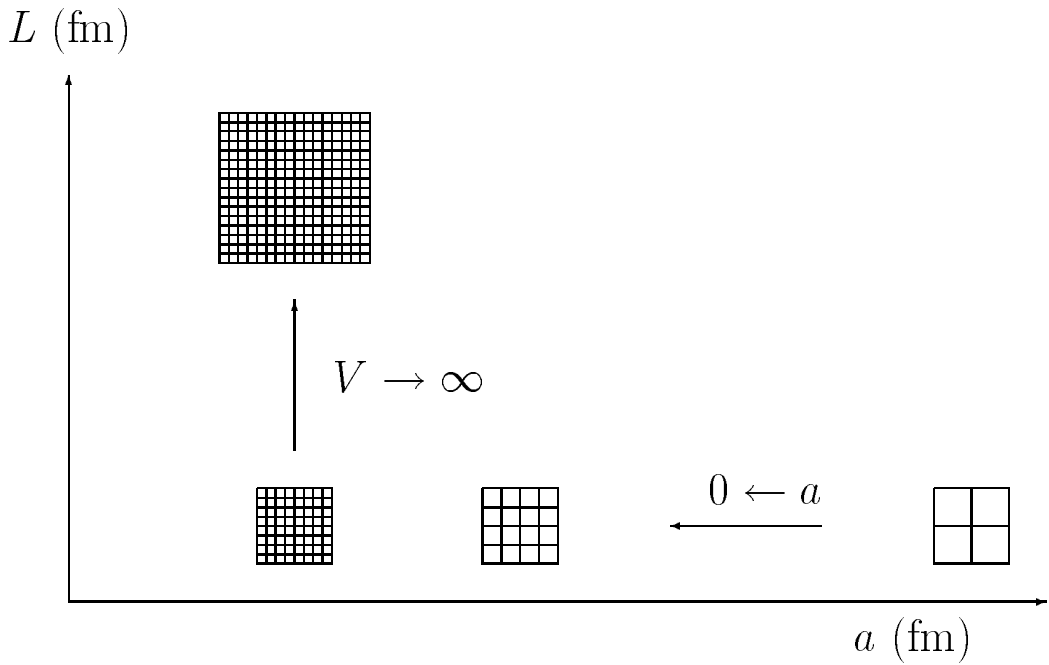

Figure 1: Illustration of the continuum and infinite-volume limits.

In practice, however, limitations in computational resources do not permit the ideal lattice QCD calculation just described. In particular, the computational cost of reducing the lattice spacing naively scales like $(L/a)^4$. (The computational cost is really higher, because of numerical problems at smaller lattice spacings.) Eq. (4) illustrates an alternative. By improving the discretization errors in the lattice action (and operators), the continuum limit can be reached at coarser lattice spacings than before. Simulations with improved actions can come at only a slightly higher computational price. The ideas underlying improvement were developed some time ago[8,9,10], and have since been revitalized[11,12,13,14].

If the quark mass is large compared to the typical QCD scale, $\Lambda_{QCD}$, effective theories are most adequate in describing the physics[15]. In that case, the lattice spacing cannot be taken to zero. Lattice-spacing errors can, however, be systematically reduced by improvement[16].

The problem is now (more or less) set up. I refer the reader again to the literature[3] for more details on the organization of typical lattice QCD calculations. I conclude this introduction with a few remarks on perturbation theory in the following subsection.

2.1. *Perturbation Theory*

Lattice QCD calculations use perturbation theory in several places:

- It guides the approach to the continuum limit.

- Short-distance quantities can be calculated non-perturbatively and compared to their perturbative expansions. It was recently shown[17] that, indeed, 1-loop perturbation theory describes most quantities considered to $3-5\%$, if a renormalized coupling like $\alpha_{\overline{MS}}$ (rather than the bare lattice coupling) is used.

- Matrix elements calculated with a lattice regulator have to be matched to their continuum counterpart by perturbation theory.

- Because quarks are confined inside hadrons, quark masses are always scheme dependent. Perturbation theory is used to convert non-perturbatively determined lattice quark masses to the perturbative continuum masses such as the pole or $\overline{MS}$ masses. Similarly, the gauge coupling can be determined non-perturbatively using lattice QCD and converted to the $\overline{MS}$ scheme at large momenta.

The lattice regulator breaks Lorentz (or Euclidean) invariance, which complicates perturbative calculations relative to those performed with Lorentz (or Euclidean) invariant regulators, such as dimensional regularization. This has prompted the development of computational techniques for higher loop perturbative calculations[18]. (Numerical) techniques for non-perturbative calculations of renormalization constants have also been developed [19,20,21]. Such techniques are very promising, because every time a new action or new operators are considered, not only must the programs be changed but also the perturbation theory has to be redone.

## 3. Quarkonium Spectroscopy

Two different formulations for fermions have been used in calculations of these spectra. In the non-relativistic limit the QCD action can be written as an expansion in powers of $v^2$ (or $1/m$), where $v$ is the velocity of the heavy quark inside the boundstate[15]; I shall henceforth refer to this approach as NRQCD. Lepage and collaborators[16] have adapted this formalism to the lattice regulator. Several groups have performed numerical calculations of quarkonia in this approach. In Refs. [22,23] the NRQCD action is used to calculate the $b\bar{b}$ and $c\bar{c}$ spectra, including terms up to $\mathcal{O}(mv^4)$ and $\mathcal{O}(a^2)$. In addition to calculations in the quenched approximation, this group is also using gauge configurations that include 2 flavors of sea quarks with mass $m_q \sim \frac{1}{2} m_s$ to calculate the $b\bar{b}$ spectrum[24,25]. The leading order NRQCD action is used in Ref. [26] for a calculation of the $b\bar{b}$ spectrum in the quenched approximation.

The Fermilab group[12] developed a generalization of previous approaches, which encompasses the non-relativistic limit for heavy quarks as well as Wilson's relativistic action for light quarks. Lattice-Spacing artifacts are analyzed for quarks with arbitrary mass. Ref. [27] uses this approach to calculate the $b\bar{b}$ and $c\bar{c}$ spectra in the quenched approximation. We considered the effect of reducing lattice-spacing errors from $\mathcal{O}(a)$ to $\mathcal{O}(a^2)$.

All but one group use gauge configurations generated with the Wilson action leaving $\mathcal{O}(a^2)$ lattice-spacing errors in the results. The lattice spacings, in this case, are in the range $a \simeq 0.05 - 0.2$ fm. Ref. [14] uses an improved gauge action together with a non-relativistic quark action improved to the same order (but without spin-dependent terms) on coarse ($a \simeq 0.4 - 0.24$ fm) lattices. The results for the $b\bar{b}$ and $c\bar{c}$ spectra from all groups are summarized in Figures 2 and 3.

The agreement between the experimentally-observed spectrum and lattice QCD calculations is impressive. As indicated in the preceding paragraphs, the lattice artifacts are different for all groups. Figures 2 and 3 therefore emphasize the level of control over systematic errors.

The first results with 2 flavors of degenerate sea quarks have appeared[25,28,29] with lattice-spacing and finite-volume errors similar to the quenched calculations, significantly reducing this systematic error. However, several systematic effects associated with the inclusion of sea quarks still have to be studied. They include the dependence on the quarkonium spectrum of the number of flavors of sea quarks and the sea-quark action (staggered vs. Wilson). The inclusion of sea quarks with realistic light-quark masses is very difficult. However, quarkonia are expected to depend only very mildly on the masses of the light quarks. This systematic error has not been included yet and should be checked numerically.

The comparison of the experimentally measured quarkonia spectra with the theo-retical calculations can be used to extract the associated Standard Model parameters: the strong coupling, $\alpha_s$, and the heavy quark masses. This is discussed in the following two sections.

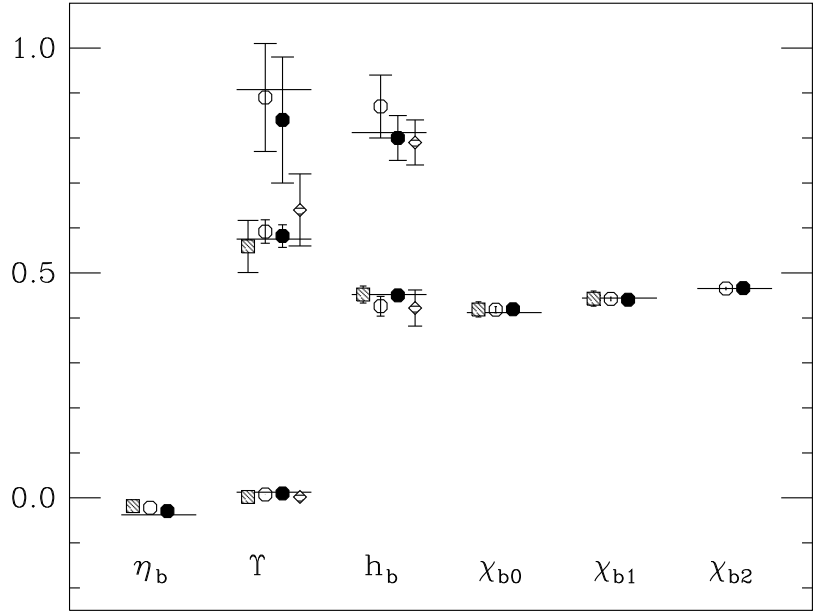

Figure 2: A comparison of lattice QCD results for the $b\bar{b}$ spectrum (statistical errors only). -: Experiment; □: FNAL [27]; ○: NRQCD ($n_f = 0$) [22]; •: NRQCD ($n_f = 2$) [24]; ◇: UK(NR)QCD [26].

## 4. The Strong Coupling from Quarkonia

At present, the QCD coupling, $\alpha_s$, is determined from many different experiments, performed at energies ranging from a few to 91 GeV[30]. In most cases perturbation theory is used to extract $\alpha_s$ from the experimental information. Experimental and theoretical progress over the last few years has made these determinations increasingly precise. However, all determinations, including those based on lattice QCD, rely on phenomenologically-estimated corrections and uncertainties from non-perturbative effects. These effects will eventually (or already do) limit the accuracy of the coupling constant determination. When lattice QCD is used the limiting uncertainty comes from the (total or partial) omission of sea quarks in numerical simulations. The determination of the strong coupling, $\alpha_s$, proceeds in three steps, outlined in the next three subsections.

*4.1. Determination of the Lattice Spacing*

The experimental input to the strong coupling determination is a mass or mass splitting, from which by comparison with the corresponding lattice quantity the lattice spacing, $a$, is determined in physical units. For this purpose, one should identify quantities that are insensitive to lattice errors. In quarkonia, spin-averaged splittings are good candidates. The experimentally observed 1P-1S and 2S-1S splittings depend only mildly on the quark mass (for masses between $m_b$ and $m_c$). Figure 4 shows the observed mass dependence of the 1P-1S splitting in a lattice QCD calculation. The comparison between results from different lattice actions illustrates that higher-order lattice-spacing errors for these splittings are small[25,27]. Figure 5 shows, in contrast, the strong dependence of the hyperfine splitting on the mass and the lattice action.

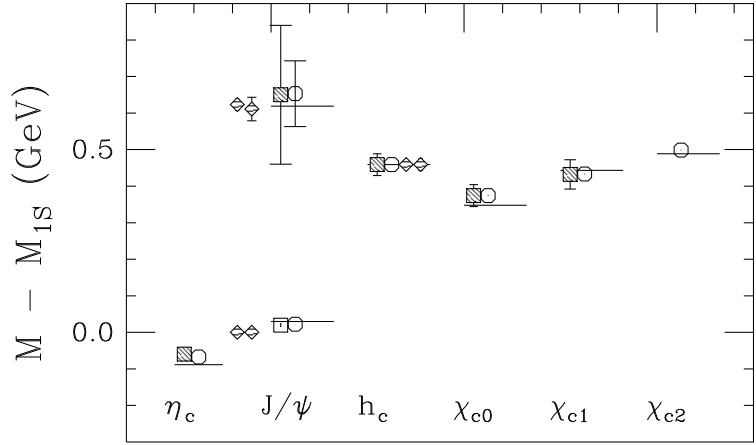

Figure 3: A comparison of lattice QCD results for the $c\bar{c}$ spectrum (statistical errors only). -: Experiment; □: FNAL [27]; ○: NRQCD ($n_f = 0$) [23]; ◇: ADHLM [14].

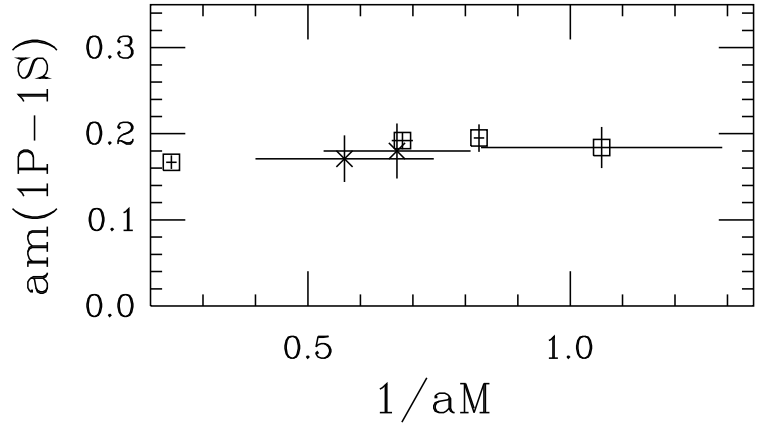

Figure 4: The 1P-1S splitting as a function of the 1S mass (statistical errors only) from Ref. [27]; □: $\mathcal{O}(a^2)$ errors; ×: $\mathcal{O}(a)$ errors.

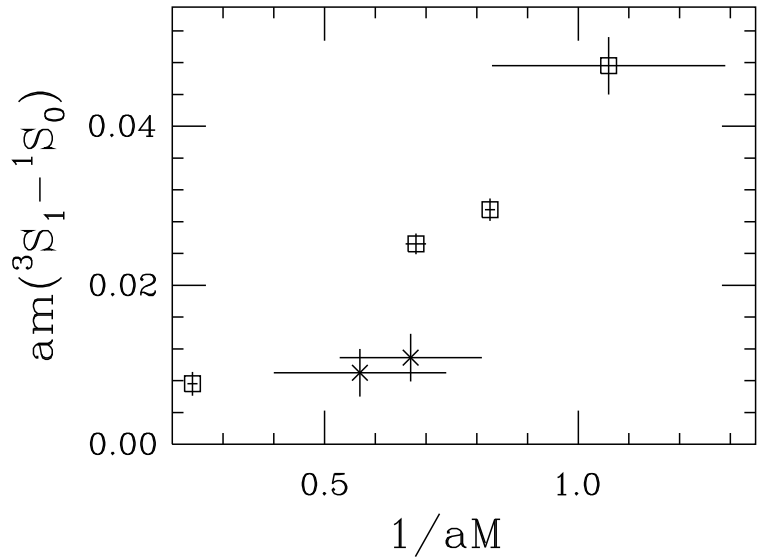

Figure 5: The hyperfine splitting as a function of the 1S mass (statistical errors only) from Ref. [27]; □: $\mathcal{O}(a^2)$ errors; ×: $\mathcal{O}(a)$ errors.

### 4.2. Definition of a Renormalized Coupling

Within the framework of lattice QCD the conversion from the bare to a renormalized coupling can, in principle, be made non-perturbatively. In the definition of a renormalized coupling, systematic uncertainties should be controllable, and at short distances, its (perturbative) relation to other conventional definitions calculable. For example, a renormalized coupling can be defined from the non-perturbatively computed heavy-quark potential[31] ($\alpha_V$). In Ref. [32] a renormalized coupling is defined non-perturbatively through the Schrödinger functional. The authors compute the evolution of the coupling non-perturbatively using a finite size scaling technique, which allows them to vary the momentum scales by an order of magnitude. The strong coupling can also be computed from the three-gluon vertex, suitably defined on the lattice[33].

An alternative is to define a renormalized coupling through short distance lattice quantities, like small Wilson loops or Creutz ratios. For example, the coupling defined from the plaquette, $\alpha_P = -3 \ln \langle \text{Tr}\, U_P \rangle / 4\pi$, can be expressed in terms of $\alpha_V$ (or $\alpha_{\overline{\text{MS}}}$) by[17]:

$$\alpha_P = \alpha_V(q)[1 - 1.19\alpha_V(q) + \mathcal{O}(\alpha_V^2)] \qquad (6)$$

at $q = 3.41/a$, close to the ultraviolet cut-off. $\alpha_V$ is related to the more commonly

used $\overline{MS}$ coupling by

$$\alpha_{\overline{MS}}(Q) = \alpha_V(e^{5/6}Q)(1 + \frac{2}{\pi}\alpha_V + \ldots) \quad . \tag{7}$$

The size of higher-order corrections associated with the above defined coupling constants can be tested by comparing perturbative predictions for short-distance lattice quantities with non-perturbative results[17]. This is consistent with the comparison of the non-perturbative coupling from Ref. [32] to perturbative predictions for this coupling using Eq. (6).

In Ref. [25] the next-to-next-to-leading order corrections to Eq. (6) have been calculated numerically from the observed deviations (from 1-loop perturbation theory) in small Wilson loops and Creutz ratios (up to size 3) at several very small lattice spacings. The dominant perturbative error then comes from the conversion to the $\overline{MS}$ coupling, which is only known to 1-loop.

The relation of the plaquette coupling in Eq. (6) to the $\overline{MS}$ coupling has recently been calculated to 2-loops[34] in the quenched approximation (no sea quarks, $n_f = 0$). The extension to $n_f \neq 0$ will significantly reduce the uncertainty due to the use of perturbation theory.

### 4.3. Sea Quark Effects

Calculations that properly include all sea-quark effects do not yet exist. If we want to make contact with the "real world", these effects have to be estimated phenomenologically or extrapolated away.

The phenomenological correction necessary to account for the sea-quark effects omitted in calculations of quarkonia that use the quenched approximation gives rise to the dominant systematic error in this calculation[35,36]. Similar ideas were used to correct for sea-quark effects in early calculations of quarkonia spectra from the heavy-quark potential calculated in quenched lattice QCD[37].

By demanding that, say, the spin-averaged 1P-1S splitting calculated on the lattice reproduce the experimentally observed one (which sets the lattice spacing, $a^{-1}$, in physical units), the effective coupling of the quenched potential is in effect matched to the coupling of the effective 3 flavor potential at the typical momentum scale of the quarkonium states in question. The difference in the evolution of the zero flavor and 3,4 flavor couplings from the effective low-energy scale to the ultraviolet cut-off, where $\alpha_s$ is determined, is the perturbative estimate of the correction.

For comparison with other determinations of $\alpha_s$, the $\overline{MS}$ coupling can be evolved to the Z mass scale. An average[30] of Refs. [35,36] yields for $\alpha_s$ from calculations in the quenched approximation:

$$\alpha_{\overline{MS}}^{(5)}(m_Z) = 0.110 \pm 0.006 \quad . \tag{8}$$

The phenomenological correction described in the previous paragraph has been tested from first principles in Ref. [28]. The 2-loop evolution of $n_f = 0$ and $n_f = 2$ $\overline{\text{MS}}$ couplings – extracted from calculations of the $c\bar{c}$ spectrum using the Wilson action in the quenched approximation and with 2 flavors of sea quarks respectively – to the low-energy scale gives consistent results. After correcting the 2 flavor result to $n_f = 3$ in the same manner as before and evolving $\alpha_{\overline{\text{MS}}}$ to the Z mass, Ref. [28] finds

$$\alpha_{\overline{\text{MS}}}^{(5)}(m_Z) = 0.111 \pm 0.005 \tag{9}$$

in good agreement with the previous result in Eq. (8). The total error is now dominated by the rather large statistical errors and the perturbative uncertainty.

Ref. [25] used results for $\alpha_s$ from the $b\bar{b}$ spectrum with 0 and 2 flavors of sea quarks to extrapolate the inverse coupling to the physical 3 flavor case directly at the ultraviolet momentum, $q = 3.41/a$. They obtain a result consistent with the old procedure, but with smaller errors:

$$\alpha_V^{(3)}(8.2\,\text{GeV}) = 0.196 \pm 0.003 \quad . \tag{10}$$

The error is dominated by the (small) statistical errors, not the extrapolation (in $n_f$) errors. The conversion to $\overline{\text{MS}}$ and evolution to the Z mass then gives:

$$\alpha_{\overline{\text{MS}}}^{(5)}(m_Z) = 0.115 \pm 0.002 \quad , \tag{11}$$

with an error now dominated by the unknown higher orders in eq. (7). A similar analysis is performed in Ref. [29] on the same gauge configurations but using the Wilson action for a calculation of the $c\bar{c}$ spectrum. The result for the coupling is consistent with Refs. [25,28].

The claimed result in Eq. (10) (or Eq. (11)) is the most accurate determination of the strong coupling constant to date. In order to confirm this result, it is desirable that the $b\bar{b}$ and $c\bar{c}$ spectra be calculated with heavy-quark actions based on Ref. [12] with the same level of statistical precision, and care with respect to systematic errors, as was done in Ref. [25]. Apart from this, the systematic errors associated with the inclusion of sea quarks into the simulation have to be checked, as outlined in section 3.

Phenomenological corrections are a necessary evil that enter most coupling constant determinations. In contrast, lattice QCD calculations with complete control over systematic errors will yield truly first-principles determinations of $\alpha_s$ from the experimentally observed hadron spectrum.

At present, determinations of $\alpha_s$ from the experimentally measured quarkonia spectra using lattice QCD are comparable in reliability and accuracy to other determinations based on perturbative QCD from high energy experiments. They are therefore part of the 1994 world average for $\alpha_s$[30]. The phenomenological corrections for the most important sources of systematic errors in lattice QCD calculations of quarkonia are now being replaced by first principles, which will significantly increase the accuracy of $\alpha_s$ determinations from quarkonia.

In a few years time, the world average for the strong coupling will be dominated by determinations of $\alpha_s$ using lattice QCD.

## 5. The Heavy Quark Masses

Because of confinement, the quark masses cannot be measured directly, but have to be inferred from experimental measurements of hadron masses, and depend on the calculational scheme employed.

In lattice QCD quark masses are determined non-perturbatively, by tuning the bare lattice quark mass $(m_Q^{\text{lat}})$ so that, for example, the experimentally observed $J/\psi$ (or $\Upsilon$) mass is reproduced by the calculation. Phenomenologically useful quark masses are the perturbatively defined pole and $\overline{\text{MS}}$ masses, which the bare lattice mass can be related to by (1-loop) perturbation theory:

$$m_Q^{\text{pole}} = Z_m^{\text{pole}} m_Q^{\text{lat}} \quad , \qquad m_Q^{\overline{\text{MS}}}(m_Q) = Z_m^{\overline{\text{MS}}} m_Q^{\text{lat}} \quad . \tag{12}$$

The heavy-quark pole mass can be determined alternatively from a calculation of the binding energy, $E_{\text{bind}}$. The ground-state energy need not equal the mass of a non-relativistic system. The binding energy can be obtained by subtracting the perturbatively calculable heavy-quark rest energy from the ground-state energy. The pole mass is then:

$$m_Q^{\text{pole}} = \frac{1}{2}(M_{Q\bar{Q}}^{\text{exp}} - E_{\text{bind}}) \tag{13}$$

This method is insensitive to errors in tuning the bare mass, because the binding energy depends only mildly on the quark mass.

Of course, as always, all systematic errors arising from the lattice QCD calculation need to be under control for a phenomenologically interesting result; in particular, the systematic error introduced by the (partial) omission of sea quarks has to be removed. The short-distance corrections that introduced the dominant uncertainty to the $\alpha_s$ determination from quarkonia are absent for the pole mass determination, because this effective mass does not run for momenta below its mass.

Ref. [24] used both methods described above for a determination of the $b$ quark pole mass from a lattice QCD calculation of the $b\bar{b}$ spectrum. As expected, a comparison of their results with zero and 2 flavors of sea quarks finds compatible results for the pole mass:

$$m_b^{\text{pole}} = (5.0 \pm 0.2) \text{ GeV} \tag{14}$$

For the $\overline{\text{MS}}$ mass, Ref. [24] quotes $m_b^{\overline{\text{MS}}}(m_b) = 4.0(1)$ GeV. The error in both results is dominated by perturbation theory.

A similar analysis is being performed in Ref. [38] for the charm and bottom quark masses from the charmonium and bottomonium spectrum. A preliminary result is $m_c^{\text{pole}} = 1.5(2)$ GeV.

The $\overline{\text{MS}}$ mass for the charm quark has also been determined from a compilation of $D$ meson calculations in the quenched approximation[39], with $m_c^{\overline{\text{MS}}}(2\,\text{GeV}) = 1.47(28)$ GeV. The error includes statistical errors from the original calculations and the perturbative error. However sea-quark effects cannot, in this case, be estimated phenomenologically, leaving this systematic error uncontrolled.

## 6. Conclusions

Quarkonia were, upon their discovery, called the hydrogen atoms of particle physics. Their non-relativistic nature justified the use of potential models, which gave a nice, phenomenological understanding of these systems. This phenomenology is at present useful to control systematic errors in lattice QCD calculations of $b\bar{b}$ and $c\bar{c}$ spectra. However, we are quickly moving towards truly first-principles calculations of quarkonia using lattice QCD, thereby testing QCD non-perturbatively. In this sense, quarkonia are still the hydrogen atoms of particle physics. Precise determinations of the Standard Model parameters $\alpha_s$, $m_b$, $m_c$, are by-products of this work.

Still lacking for a first-principles result is the proper inclusion of sea quarks. The most difficult problem in this context is the inclusion of sea quarks with physical light quark masses. At present, this can only be achieved by extrapolation (from $m_q \simeq 0.3 - 0.5 m_s$ to $m_{u,d}$). If the light quark mass dependence of the quarkonia spectra is mild, as anticipated, the associated systematic error can be controlled. First-principles calculations of quarkonia could then be performed with currently available computational resources.

## 7. Acknowledgements

I thank the organizers for an enjoyable conference and E. Eichten, A. Kronfeld, P. Lepage, P. Mackenzie, and J. Shigemitsu for discussions while preparing this talk.